\documentclass[prl,aps,twocolumn,tightenlines,floatfix,nofootinbib,showpacs]{revtex4-1}

\usepackage{epsf}    
\usepackage{graphicx}
\usepackage{epstopdf}

\usepackage{color}
\usepackage{amsmath}
\usepackage{graphicx}
\usepackage{amssymb}
\usepackage{psfrag}

\newcommand{\beq}{\begin{equation}}
\newcommand{\eeq}{\end{equation}}
\newcommand{\bay}{\begin{array}}
\newcommand{\eay}{\end{array}}
\newcommand{\beqy}{\begin{eqnarray}}
\newcommand{\eeqy}{\end{eqnarray}}

\newcommand{\rmd}{\mathrm{d}}
\newcommand{\brac}[1]{\left({#1}\right)}

\newcommand{\pd}[2]{\frac{\partial{#1}}{\partial{#2}}}
\newcommand{\td}[2]{\frac{\rmd{#1}}{\rmd{#2}}}
\newcommand{\curl}{\nabla\times}

\newcommand{\bB}{{\bf B}}
\newcommand{\be}{{\bf e}}

\newcommand{\fmag}{\boldsymbol{\mathfrak{F}}_{\mathrm{mag}}}

\newcommand{\eps}{\epsilon}

\renewcommand{\rmp}{\mathrm{p}}
\newcommand{\rmn}{\mathrm{n}}
\newcommand{\rmx}{\mathrm{x}}

\newcommand{\mux}{\tilde{\mu}_\rmx}

\newcommand{\rhop}{\rho_\rmp}

\begin{document}

\title{Magnetic fields in superconducting neutron stars}

\author{S. K. Lander}
\email{samuel.lander@uni-tuebingen.de}
\address{Theoretical Astrophysics, IAAT, University of T\"ubingen, 72076 T\"ubingen, Germany}

\begin{abstract}
The interior of a neutron star is likely to be predominantly a mixture of
superfluid neutrons and superconducting protons. This results in the
quantisation of the star's magnetic field into an array of thin
fluxtubes, producing a macroscopic force very different from the Lorentz force of normal
matter. We show that in an axisymmetric superconducting equilibrium the behaviour
of a magnetic field is governed by a single differential
equation. Solving this, we present the first self-consistent
superconducting neutron star equilibria with poloidal and mixed poloidal-toroidal
fields, also giving the first quantitative results for the
corresponding magnetically-induced distortions to the star. The
poloidal component is dominant in all our configurations. We suggest
that the transition from normal to superconducting matter in a young
neutron star may cause a large-scale field rearrangement.
\end{abstract}


\maketitle


\emph{Introduction}.---There is now compelling evidence that the
bulk of a neutron star is composed of superfluid neutrons and
superconducting protons. It has long been thought that neutron stars
would be cold enough to contain these states of
matter\citep{migdal,baym_pp,sauls_review}, based on
expectations from the theory of terrestrial
superconductivity\citep{BCS}. A long-standing piece of evidence in
favour of superfluidity is the phenomenon of pulsar glitches ---
sudden events in which the star spins \emph{up} slightly \citep{and_itoh}.
The best explanation for the larger glitches is as a transfer of
angular momentum from the more rapidly rotating superfluid to the crust.
In addition, recent observations of the rapidly cooling young neutron star in
the Cassiopeia A supernova remnant are well explained by the onset of
neutron superfluidity together with proton superconductivity in the core
\citep{page_letter,shternin_letter}, suggesting that these components
will be present in a neutron star below a critical temperature of
around $10^9$ K. The most highly magnetised neutron stars, the magnetars, have high observed
temperatures which might suggest their transition to superfluidity
is delayed. This is due to crustal heating however, whilst the
thermally decoupled magnetar core is expected to cool rapidly below the
critical temperature \citep{ho_glamand}. Thus
\emph{all} observed neutron stars are likely to contain superfluid and
superconducting components.

In terrestrial superconductors the Meissner effect expels any magnetic field
below some critical strength, but in neutron stars the expulsion
timescale is extremely long \citep{baym_pp}, so the magnetic field
exists in a metastable state. Nonetheless it is affected: in a large
part of the star the protons are expected to form a type-II
superconductor, meaning that the field will be quantised into fluxtubes
surrounded by unmagnetised matter.  On the
macroscopic scale this changes the nature of the magnetic force from
the Lorentz force of normal matter to a fluxtube tension force \citep{easson_peth,mendell,GAS}. The
innermost part of the core may
exhibit type-I superconductivity\citep{sedrakian,jones06}, with
large regions of alternating normal and superconducting matter;
the effect of this on the global magnetic field is unknown at present.  Alternatively, the star's hadronic matter could give way to
an inner core of `colour superconducting' quark matter\cite{alcock,GJS}.

A neutron star's magnetic field can play a variety of important
roles. It affects the temperature and rotational evolution of the star
--- and rotation is the key observational feature used to determine
the star's age. Understanding the interior fields of apparently different
classes of neutron star could help unify them into a single canonical stellar model
\citep{perna_pons,kaspi}. Magnetic field effects could explain the differing nature
of glitches in pulsars and magnetars and post-glitch recovery\citep{easson_glitch}. In magnetars, the
magnetic field provides the energy that powers their giant flares and is important for
understanding their observed quasi-periodic
oscillations\citep{israel,stroh_watts}. Finally, a magnetic field
induces a distortion which will generally not be aligned with the star's rotation axis. This
system will therefore produce gravitational waves\citep{bon_gour},
perhaps at an amplitude great enough for future detection.

Motivated by the above reasons, there has been a great deal of recent
work on neutron star magnetic fields, but almost all of it based on
models assuming normally-conducting matter. This may be partly because
this case is more familiar, thanks to the large body of literature on
non-degenerate stars \citep{mestel_book}. For neutron stars
superconductivity is a key missing ingredient, whose inclusion is an
essential step towards more realistic models.

In this letter we try to lay some foundations for the modelling of
superconducting neutron stars. We study equilibrium configurations,
motivated by the observation that neutron star magnetic fields appear to be
long-lived, evolving only on long timescales. We show that in axisymmetry the magnetic
field of a superconducting star is governed by a single differential equation. This is analogous to the
`Grad-Shafranov' equation for normal matter\citep{grad_rubin,shafranov}, 
but more complicated: it involves
terms related to the local field strength. We 
describe a method of solution for the equation and present the
results. Other than the special case of a purely toroidal field\citep{akgun_wass,LAG}, these
are the first self-consistent solutions for magnetic fields in a
superconducting star (see, however, the simplified poloidal-field
model constructed in \citep{roberts})\footnote{Since the submission of
  this letter, a new study on poloidal fields has appeared \citep{hen_wass}}. We
compare with the corresponding normal-matter results and discuss the
implications for the internal magnetic fields of neutron stars.


\emph{Axisymmetric magnetic stars}.---We model a neutron star as a
three-fluid system, with superfluid neutrons, electrons and type-II
superconducting protons. The electrons have negligible inertia however, and
their chemical potential can be incorporated into that of the
protons. Therefore we can reduce to a two-fluid system of
equations, denoting neutron quantities with a subscript `n' and
the combined proton and electron quantities with a `p'. We choose a
rather idealised equation of state, a two-fluid analogue of  
a polytrope \citep{prix_ca}, so that the chemical potential $\mux$ of
each species ($\rmx=\{\rmn,\rmp\}$) is a function of the corresponding mass density
$\rho_\rmx$: $\mux=\mux(\rho_\rmx)$. Working in cylindrical polar
coordinates $(\varpi,\phi,z)$, the two Euler equations for our
model may be written:
\beq \label{x_Euler}
\nabla\brac{\mux+\Phi-\frac{\varpi^2\Omega_\rmx}{2}}=\frac{{\bf F}_\rmx}{\rho_\rmx},
\eeq
where $\Phi$ denotes gravitational potential,
$\Omega_\rmx$ rotation rate and ${\bf F}_\rmx$ magnetic force. 
Although we only consider non-rotating models here, the following
derivations and numerics follow through for cases with corotating
neutrons and protons, $\Omega_\rmn=\Omega_\rmp$. In general one would
expect an \emph{entrainment} effect, leading to coupling between
neutrons and protons and to an effective magnetic force on the
neutrons. We neglect this for simplicity, so that
${\bf F}_\rmn=0$, and write the proton force as $\fmag$.
The particle species are therefore only coupled through gravity:
\beq
\nabla^2\Phi = 4\pi G(\rho_\rmn+\rho_\rmp).
\eeq
Now, taking the curl of the proton Euler equation, 
we see that there exists a scalar $M$ such that
\beq
\frac{\fmag}{\rho_\rmp}=\nabla M,
\eeq
regardless of whether the protons are normal or superconducting.
Another universal result is that $\bB$ must be divergence-free; using
this together with the assumption of axisymmetry allows us to write
\beq \label{streamfn}
\bB = \frac{1}{\varpi}\nabla u\times\be_\phi + B_\phi\be_\phi,
\eeq
which defines the streamfunction $u$. Note that $\bB\cdot\nabla u=0$;
field lines are parallel to constant-$u$ contours.

\emph{Normal matter}.---In this familiar case $\fmag$ is the Lorentz force, given by
\beq
\fmag=\frac{1}{4\pi}(\curl\bB)\times\bB.
\eeq
It can be shown that the toroidal field component is governed by a
function $f_N(u)=\varpi B_\phi$ \citep{grad_rubin,shafranov}. Using
this and the general results of the last section, one 
may derive the \emph{Grad-Shafranov equation}:
\beq \label{grad_shaf}
\Delta_*u
 \equiv \pd{^2u}{\varpi^2}-\frac{1}{\varpi}\pd{u}{\varpi}+\pd{^2u}{z^2}
 = -4\pi\rhop\varpi^2\td{M}{u} - f_N\td{f_N}{u}.
\eeq
This governs the form of a poloidal or mixed poloidal-toroidal field
in axisymmetric equilibrium. The single-fluid version of this has been
the basis of numerous studies of magnetic stars.


\emph{Superconducting matter}.---By averaging the contribution of the
fluxtubes in a type-II superconductor, one arrives at a macroscopic expression for the
magnetic force\citep{easson_peth,mendell,GAS}:
\beq \label{orig_fmag}
\fmag = -\frac{1}{4\pi}\Bigg[
                  \bB\times(\curl{\bf H}_{c1})
                   +\rhop\nabla\brac{B\pd{H_{c1}}{\rho_\rmp}}
                                  \Bigg],
\eeq
where ${\bf H}_{c1}=H_{c1}\hat\bB$ is the `microscopic' critical
field, $\bB$ the smooth-averaged `macroscopic' field and $\hat\bB=\bB/B$ the unit
tangent vector to the magnetic field. In the absence of
entrainment $H_{c1}=h_c\rho_\rmp$ to a good approximation, where $h_c$ is a constant\citep{GAS}. Using
this relation and equations \eqref{streamfn} and
\eqref{orig_fmag}, we get an expression for the magnetic force with various poloidal terms (see
\citep{LAG}) and one toroidal term, which must be zero by axisymmetry:
\beq \label{parallelnablas}
(\mathfrak{F}_{\mathrm{mag}})_\phi=\nabla u\times\nabla(\rho_\rmp\varpi\hat{B}_\phi)=0.
\eeq
This gives us a dichotomy:
satisfying \eqref{parallelnablas} leads either to a
mixed poloidal-toroidal field or a purely toroidal field. For the
latter case, we satisfy \eqref{parallelnablas}
by taking $\nabla u={\bf 0}$; this has been discussed in earlier work
\citep{akgun_wass,LAG}.  For the former,
we require that $\nabla u$ and $\nabla(\rho_\rmp\varpi\hat{B}_\phi)$
be parallel, which leads to
\beq \label{f_sc}
\rho_\rmp\varpi\hat{B}_\phi = f(u)
\eeq
for some function $f$. In the special case $f(u)=0$ the field is
purely poloidal.

One key step in the derivation of the Grad-Shafranov equation is showing that
$M=M(u)$. In the superconducting case this is no longer true. We can, however, define a related
function $y$ which \emph{is} a function of $u$:
\beq \label{y_defn}
y(u) = \frac{4\pi M}{h_c}+B.
\eeq
Using the functions $y(u)$ and $f(u)$, together with the poloidal
magnetic-force terms from \citep{LAG}, we arrive at a single
differential equation governing the magnetic field:
\beq \label{supercon_GS}
\Delta_* u 
  = \frac{\nabla\Pi\cdot\nabla u}{\Pi}-\varpi^2\rhop\Pi\td{y}{u}-\Pi^2 f\td{f}{u}\ ,       
\eeq
where $\Pi\equiv B/\rhop$. The above equation is the
equivalent of the Grad-Shafranov equation when the protons form a type-II
superconductor instead of being a normal fluid. The result for a
single-fluid superconducting star may be obtained by replacing $\rhop$
with $\rho$, the total density, and is valid for barotropic
equations of state. The most significant differences from the normal case 
are the presence of the $\Pi$ factors and the fact
that the magnetic force no longer appears explicitly through the function $M$. 


\emph{Superconducting core and normal `crust'}.---The superfluid and
superconducting matter of a neutron star's core does not extend to the
stellar surface; instead the star has an elastic crust of normal
matter. It is also numerically difficult to solve equation
\eqref{supercon_GS} matched directly to a potential field
($\curl\bB=0$) exterior. For these reasons we choose a canonical
neutron star model with a core of superfluid neutrons and type-II
superconducting protons, matched to a single-fluid relaxed `crust' composed
solely of normal protons; this in turn is matched to a vacuum exterior with
potential field at the stellar surface $R_*$.

The complicated physics at a neutron star's crust-core boundary may
well include the presence of a current sheet and a discontinuity in the
magnetic field (\citep{hen_wass} contains some discussion of this issue). In
addition, the pinning of fluxtubes to the crust could be
important. Quantifying these effects is beyond the scope of this 
paper, however, so for this first study we assume continuity of
the magnetic field at the boundary. We also demand magnetic force
balance, matching the smooth-averaged fluxtube tension force and the
Lorentz force at the crust-core interface. More specifically, we define the crust-core
boundary as an isopycnic contour $\rhop=\rhop^{cc}$ (at a
near-constant radius of $r=0.9R_*$), requiring that the
magnetic-force scalar function $M$ be continuous and that
${\bf H}_{c1}\to\bB$ there. This may be
done by defining the superconducting functions $y(u),f(u)$ in terms of
their normal-matter counterparts $M_N(u),f_N(u)$:
\beq
y(u) = h_c\rhop^{cc}+4\pi M_N(u)/h_c,
\eeq
\beq
f(u) = f_N(u)/h_c.
\eeq
This then
allows for smooth matching of the right-hand sides of the two
governing equations \eqref{grad_shaf} and \eqref{supercon_GS}.


\emph{Numerics}.---The governing equation for superconducting
equilibria \eqref{supercon_GS} is harder to solve than the
Grad-Shafranov equation \eqref{grad_shaf}. The latter
is itself unusual in having the argument $u$ appear on
both left- and right-hand sides, but the superconducting version also
has the quantity $\Pi$, which implicitly involves derivatives of $u$:
\beq
\Pi=\frac{|\nabla u|}{\sqrt{\varpi^2\rhop^2-f^2}}.
\eeq
With direct solution seeming infeasible, this problem is suited to a numerical
iterative method, where both left- and right-hand sides are
gradually updated until the scheme converges and produces a consistent
solution for $u$. We use an adapted self-consistent
field method \citep{hachisu,LAG}, allowing for high multipolar structure, and employ an underrelaxation step
for the solution of \eqref{supercon_GS}.

Using the virial theorem \citep{chandra} we confirm that our results are indeed
equilibrium solutions, with a relative error of the order
$10^{-5}$. This error decreases accordingly with increasing resolution. As
another check, we have constructed equilibria with increasing
field strength (and hence distortion), confirming that the induced
distortion scales in the expected manner: linearly in $H_{c1}B$
\citep{jones75,easson_peth}. Details of these tests and the numerical
method will be presented in a later paper.


\begin{figure}
\centerline{\includegraphics[width=0.5\textwidth]{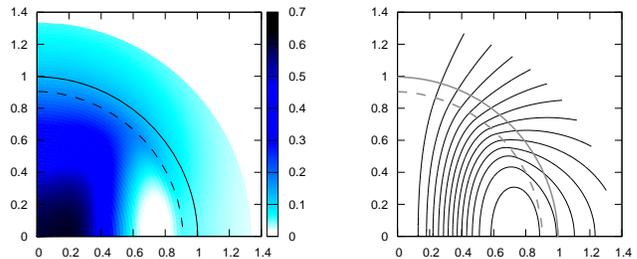}}
\caption{Structure of a poloidal magnetic field in a superconducting
  star. On the left we show the magnitude of the field and on the
  right its direction (i.e. the field lines). The stellar surface $R_*$ is
  indicated with the solid arc at a dimensionless radius of
  unity, whilst the dashed line at $0.9R_*$ shows the location of the
  crust-core boundary.
\label{pol_fig}}
\end{figure}

\emph{Results}.---We consider a stratified stellar model with a
composition gradient, i.e. the proton fraction $\rhop/\rho$ varies within the
star. Various forms for the magnetic functions are
permissible. We choose $f_N(u)=a(u-u_{\mathrm{int}})^\zeta$ when
$u>u_{\mathrm{int}}$ and $f_N(u)=0$ otherwise, where
$u_{\mathrm{int}}$ is the largest contour of $u$ (i.e. field line)
that closes within the star; this avoids having an exterior
current. We also take $M_N(u)=\kappa u^2$ unless otherwise stated. Here $a,\zeta$ and $\kappa$ are
constants related to the strength of the magnetic field
components. The figures are presented in dimensionless units, since
the structure of the magnetic field is essentially independent of its strength.

We begin by looking at a purely poloidal field, in figure \ref{pol_fig}. This is broadly
similar to the corresponding field in a normally-conducting star, with
the field strength attaining a maximum $B_{\textrm{max}}$ deep in the star and
vanishing in the centre of the closed field-line region. The main difference
is that the superconducting star shown has a far larger `weak-field'
region than a model with normal protons; more specifically, a larger
volume of the superconducting star has a field strength
$B<0.1B_{\textrm{max}}$. This effect is even more pronounced for
models with $M_N(u)=\kappa u$.

\begin{figure}
\centerline{\includegraphics[width=0.4\textwidth]{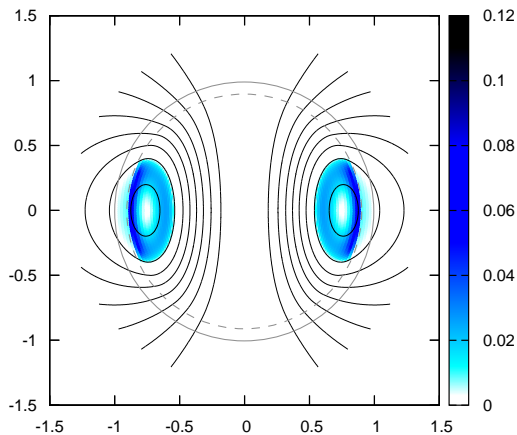}}
\caption{A mixed poloidal-toroidal magnetic field in a superconducting
  star. We show the poloidal field lines in black and the magnitude of the
  toroidal-field component with the colour scale. The stellar surface
  and crust-core boundary are indicated with the solid and
  dashed circles. Note that the toroidal component has a tubular
  structure, vanishing in the centre of the closed field line region.
\label{mix_fig}}
\end{figure}

In figure \ref{mix_fig} we show a typical mixed poloidal-toroidal
field configuration. For stars with normal protons the toroidal
component fills the weak-field region shown in figure \ref{pol_fig},
producing a `twisted-torus' configuration\citep{LAG}. This is partially true
here too --- but at the centre of the closed field line region, where the
poloidal field vanishes, the toroidal component vanishes too. The
resulting toroidal-field geometry is hence tubular (in three
dimensions). The mixed-field configuration 
in figure \ref{mix_fig}, like all those we have found, is
dominated by the poloidal component; the toroidal component only
contributes $0.7\%$ of the total magnetic energy. This may be related
to the fact that equation \eqref{supercon_GS} has a purely-poloidal
limit but no purely-toroidal one.

Next we look at the magnetically-induced ellipticity
$\eps=(Q_{\mathrm{eq}}-Q_{\mathrm{pole}})/Q_{\mathrm{eq}}$,
where $Q_{\mathrm{eq}},Q_{\mathrm{pole}}$ represent the components of
the star's quadrupole moment along the equator and pole. We rescale to
a typical $1.4$-solar mass neutron star with a radius of $10$ km and 
assume a purely poloidal field. The mass
of the normal-fluid crust is small, so the ellipticity scaling is well
approximated by that of a purely superconducting star:
\beq \label{sc_ellip}
\eps = 3.4\times 10^{-8}\brac{\frac{B_s}{10^{12}\ \textrm{G}}}
                                         \brac{\frac{H_{c1}(0)}{10^{16}\ \textrm{G}}},
\eeq
where $B_s$ denotes the surface field strength at the pole.
We adopt a central critical field of $H_{c1}(0)=10^{16}$ G, using the 
approximate formula given in \citep{GAS}. 
For comparison, in the same stratified two-fluid model but with normal
protons we find (using the code described in \citep{LAG}):
\beq \label{normal_ellip}
\eps = 2.6\times 10^{-11}\brac{\frac{B_s}{10^{12}\ \textrm{G}}}^2.
\eeq
A simple way to approximate the ellipticity of a superconducting
star is to take a result for normal matter and scale it up by a
factor $H_{c1}/B$, taking $H_{c1}=10^{15}$ G. Comparing 
our ellipticity formulae \eqref{sc_ellip} and \eqref{normal_ellip},
both from self-consistent calculations, we see that for a given $B_s$
this approach would underestimate the star's distortion by around $30\%$.
Choosing $M_N(u)$ as a higher power of $u$ increases the ratio of
average internal field $\bar{B}$ to $B_s$ and hence the ellipticity at
a fixed value of $B_s$. This is summarised in table
\ref{Mu_variation}.


\begin{table}
\caption{The ratio $\bar{B}/B_s$ of average internal field to polar
  field and the prefactor $k_\eps$ of the ellipticity formula
  \eqref{sc_ellip}, for different poloidal-field configurations
  (specifically, different relations between the streamfunction $u$
  and the magnetic-force function $M_N(u)$).}
\label{Mu_variation}
\begin{tabular}{|c|c|c|c|c|} 
\hline
$M_N(u)$ scaling &$\ \ u\ \ $&$\ \ u^2\ \ $&$\ \ u^3\ \ $&$\ \ u^4\ \ $\\
\hline
$\bar{B}/B_s$      & $1.5$ & $2.1$ & $2.7$ & $2.5$ \\
$k_\eps[10^{-8}]$ & $2.5$ & $3.4$ & $3.9$ & $3.5$ \\
\hline
\end{tabular}
\end{table}


\emph{Discussion}.---
We have described a method to solve for the
magnetic field in a neutron star with type-II superconducting
protons. The magnetic force is more complicated in
this case than for normal matter, but for an axisymmetric equilibrium
we show that it may be simplified to a single differential equation in
the streamfunction and the local field strength, in analogy with the
Grad-Shafranov equation of normal matter. We solve this using an
iterative scheme, presenting the first self-consistent models of a
superconducting neutron star (other than the special case of a purely toroidal field).

Perhaps the most notable difference from normal-matter models is
the generic appearance of a region in the star where the field
strength vanishes. This may have repercussions for 
the dynamics of a neutron star; in normal matter it is known that such
a region leads to an instability\cite{wright,mark_tay}. Whether such
an instability occurs in superconducting matter is an interesting open
question.

The interior field of a neutron star cannot necessarily be inferred
from the observed exterior dipole field. If many poloidal field lines close
within the star, or if there is a strong toroidal component, the
internal field could be much stronger than expected from outside ---
a `hidden' energy reservoir for the star. In our models the average interior field strength is
$1.5-2.7$ times that at the polar surface. The contribution of the
toroidal component appears to be generically small, however; in our
equilibria it accounts for less than $\sim 1\%$ of the magnetic
energy. This is different from recent simulations for main-sequence
stars, which found stable equilibria with \emph{large} toroidal components\cite{braith}.

A leading theory for the origin of neutron star magnetism (for
magnetars in particular) is that dynamo action in the young star
generates a strong, dominantly toroidal field\cite{thomp_dunc} --- in
contrast to our equilibria for mature neutron stars. This may cast
doubt on the validity of our equilibria or 
the dynamo scenario. Alternatively, both could be reasonable --- then, as a
hot young neutron star cools to a multifluid state with superconducting protons, its
strong toroidal field would no longer be in equilibrium. It
would have to undergo large-scale rearrangement to a
poloidal-dominated equilibrium, a potentially violent transition which
could be observable.

The effect of superconductivity on neutron star magnetic fields has been, to
date, neglected by the vast majority of studies. This letter
demonstrates that in the simplest equilibrium models it may be accounted 
for using similar techniques as for normal-matter stars. Two key
issues which future work should address are the presence of a magnetic
force on the neutrons and the physics at the crust-core
boundary. Beyond this, superconductivity will surely also play an
important role in the evolution and dynamics of neutron stars.


\acknowledgements 
I am pleased to thank Nils Andersson and Kostas Glampedakis for their
helpful comments on a draft of this paper, and Ira Wasserman for
useful correspondence. This work was supported by
the German Science Foundation (DFG) via SFB/TR7.


\end{document}